\documentclass[preprint,10.5pt]{elsarticle}
\usepackage{graphicx}
\usepackage{amssymb, amsmath}
\usepackage{fancyhdr}
\usepackage[mathscr]{eucal}

\begin{document}

\begin{frontmatter} % elsarticle
%%%
%%% Title page
%%%
\title{Dynamical Galam model}

\author[tc]{Taksu Cheon}
\ead{taksu.cheon@kochi-tech.ac.jp} % underscore _ requres \
\author[sg]{Serge Galam}
\ead{serge.galam@sciencespo.fr} % underscore _ requres \

\address[tc]{
Laboratory of Physics, Kochi University of Technology, \\ Tosa Yamada, 
Kochi 782-8502, Japan} 
\address[sg]{
CEVIPOF - Centre for Political Research, Sciences Po and CNRS,\\
98 rue de l'Universit\'e, 75007 Paris, France} 

%\date{\today}
\date{February 23, 2018}

\begin{abstract}
We introduce a model of  temporal evolution of political opinions which amounts to a dynamical extension of Galam model in which the proportions of inflexibles are treated as dynamical variables.  We find that the critical value of inflexibles in the original Galam model now turns into a fixed point of the system whose stability controls the phase trajectory of the political opinions.  The appearance of two phases is found, in which majority-preserving and regime-changing limit cycles are respectively dominant, and  the phase transition between them is observed.
\end{abstract}

%\pacs{05.10.-a, 87.23.Ge, 89.75.Fb}
%\keywords{sociophysics, opinion dynamics, critical phenomena}

\end{frontmatter} % elsarticle
%%%

\section{Introduction}
We have witnessed in past two decades, the rise of a new type of mathematical models of human society, distinct from conventional microeconomics, which focus on the formation of social and political opinions in a society \cite{CF09, GA12, BR12, FS14}.  %
These models have revealed unexpected similarity between certain characteristics of human society and the statistical properties of condensed matter systems.  We can nowadays, for example, talk about the phase transition in opinion dynamics in human society \cite{HM10, HM15, MH15}.

The Galam model is their prime example, in which heterogenous ``agent types'' played a critical role \cite{GA05a, CM16}. 
In particular, one salient trait to discriminate agents is the individual ability of an agent to eventually shift opinion from the one it had adapted earlier. An agent who skips to its initial choice is tagged as an inflexible against floaters who are susceptible of shifting opinion driven by local exchanges with other within small groups of discussion. 
%v5 (1.3
The concept of inflexibilty in opinion forming was first introduced to study group decision making using Ising spin like modeling with quenched random local field \cite{mosco}. 
 
%v5 (1.4
Incorporating into opinion dynamics with moving agents within the so-called Galam dynamics model \cite{GJ07, gem}, it leads to numerous studies  \cite{nuno1, nuno2, nuno3, rand, bollen, kasia}. 
%v5 (1,2
%v5 R3 1
Latter, other denominations have been used like ``zealots'' \cite{ mobi} and ``committed'' agents \cite{bolek}. While all these works consider the proportions of inflexibles as fixed external parameters, a study has investigated numerically the building of inflexibilty as an internal dynamics, which is a function of the number of times an agent found itself having the same opinion \cite{ martins}. Here we extend the investigation of inflexibility considering a novel feature, which reinterprets these inflexibles as opinionated determined minority, or minority with extremist views.  

%v5 R2 5
%v5 R3 1
Observing that in real world politics, a motivated minority is often found to be  the driving force in political regime change \cite{mosco-mino}.  It is natural to assume that the number of determined minority is not fixed, but can increase and decrease depending on the environment inflexibles and floaters are in.  Accordingly we make inflexibles dependent on their overall local environment. A hostile environment tends to strengthen them increasing their number while their victory tends to weaken them  decreasing their number.
 
 %v5 R3 5
In this paper we incorporate such a local dependance of inflexibles extending the Galam model of majoritarian dynamics which has the characteristic agent type ``inflexibles'' \cite{GA05a, gem, TG15} by adding update rules for the production and reduction of inflexibles. Analyzing the model both numerically and analytically with linearization around the fixed points, we found the emergence of fixed points and limit cycles.  

It turns out that there are two types of limit cycles, ones that preserve the majority, and the other that causes the cyclic alternation of winning opinions.    We show that with the change of system parameters, a phase transition-like behavior is observed between one phase where majority-conserving cycle dominates, and the other phase where only majority-alternating cycles are present. Our model being more realistic with respect to inflexibility, which is a major ingredient of real social system, it may shed new light to understand political cycles in democratic countries with elected governments.

%v5 R3 1
The paper is organized as follows.  We introduce the dynamical systems model of Galam opinion dynamics in the second section.  In the third section, we present the results of numerical calculations.  In the fourth section, we analyze the fixed points of the dynamical system, and discuss their stability with the linearized map analysis. The existence of phase transition-like behavior is also pointed out.  In the fifth section, the analysis of oscillation period is presented.  The paper is concluded with some discussions in the last, sixth section.

\section{The dynamical extension of Galam model}

We construct a dynamical extension of the Galam model, that enables describing the temporal variation of the number of inflexibles, thereby establishing a model for the secular changes of political majority.  Our basic observations on the role of  extremism in the political process are:
%
%v5 (2.1
\begin{itemize}
\item{} It is a committed few who often drive political change by tirelessly pushing their cause.
\item{} Extremists thrive in hostile environment, but lose their edge after success.
\end{itemize}
%
%v5 (2.2, (2.3
To model this tendency, we assume that, after each update in Galam model, the number of inflexible agents increase with a probability $f$ if the local majority goes against them, while it decreases with a probability $g$.   We  assume that the probabilities are proportional to the number of existing inflexibles.  We also include the appearance of inflexibles inside the group which has no inflexibles, which we represent by the probability $h$.  Here $f$, $g$ ad $h$ are positive numbers between $0$ and $1$.  

%v5 (2.4
%v5 R2 1
%v5 R3 2
Starting with $N$  agents capable of independently taking two values $1$ and $0$, signifying the support for party $A$ and $B$, respectively, we repeat the following process.   All agents are randomly divided into groups of $r$ agents, and within each group the values of agents are updated to conform to the initial local majority within the group, with the following exceptions:
\begin{itemize}
\item There are agents called {\em inflexibles}, who do not follow the group-majoritarian update rule simply keeping their own fixed value.  There are both $A$-inflexibles and $B$-inflexibles whose respective preset values are $1$ and $0$.  Agents who do not belong to inflexibles are called {\em floaters}.
%v5 (2.5
\item Within a group whose majority has gone to party $A$,  after the update the number of $A$-inflexibles decreases probabilistically by the factor $1-g$, and the number of $B$-inflexibles increases probabilistically by factor the $1+f$.  When there are no $B$-inflexible, one appears anew in the group with probability $h$.
\item Within a group whose majority has gone to party $B$,  the number of $B$-inflexibles decreases probabilistically by the factor $1-g$, and the number of $A$-inflexibles increases probabilistically by the factor $1+f$.  When there are no $A$-inflexible, one appears anew in the group with probability $h$.
\end{itemize}
Here we assume $N$ to be a multiple of $r$.   At time step $t$, we denote the ratio of agents supporting the party $A$ by $p_t$, and resultantly the ratio of agents supporting the party $B$ by $1-p_t$.  It includes both floaters and inflexibles. The ratio of $A$-inflexibles and $B$-inflexibles at the time step $t$ are denoted by $a_t$ and $b_t$ respectively.
%
%--------------------
\begin{table}[ht]%[htbp]
  \centering
  %\topcaption{Table captions are better up top} % requires the topcapt package
  \begin{tabular}{@{} lccccccccc @{}} % Column formatting, @{} suppresses leading/trailing space
   \hline % \toprule
%     \multicolumn{2}{c}{\qquad\ \  b : 0 start} \\
%     \cmidrule(r){2-3} % Partial rule. (r) trims the line a little bit on the right; (l) & (lr) also possible
    $k$ & agents & update   &  $m_k$ &  $P_k$ & $K_k^{f1} $ & $K_k^{f0} $ & $K_k^{a} $ & $K_k^{b} $ \ \ \\
     [0.7ex]
     \hline \\% \midrule
     [-2.5ex]
1&  000&{\small 000} $\rightarrow$ {\small 000} &1& $d^3$ & $0$ & $\frac{3-h}{3}$ & $\frac{h}{3}$ & $0$ \\
      [1.0ex]
2& 100&{\small 100} $\rightarrow$ {\small 000} & 3 & $ u d^2 $  & 0 & 1 & 0 & 0  \\
3& 110&{\small 110} $\rightarrow$ {\small 111} &  3 & $ u^2 d $  & 1 & 0 & 0 & 0  \\
4& 111&{\small 111} $\rightarrow$ {\small 111} &  1 & $ u^3 $  
     & $\frac{3-h}{3}$ & 0 & 0 & $\frac{h}{3}$ \\
     [0.7ex]
     \cline{2-3} \\% \cmidrule(l){2-3}\\
     [-2.5ex]
5& b00&{\small b00} $\rightarrow$ {\small b00} &  3 & $ b d^2 $  
     & 0 & $\frac{2+g}{3}$ & 0 & $\frac{1-g}{3}$  \\
6& b10&{\small b10} $\rightarrow$ {\small b00} &  6 & $ b u d $  
     & 0 & $\frac{2+g}{3}$ & 0 & $\frac{1-g}{3}$  \\
7& b11&{\small b11} $\rightarrow$ {\small b11} &  3 & $ b u^2 $  
     & $\frac{2-f}{3}$ & 0 & 0 & $\frac{1+f}{3}$   \\
     [1.0ex]
     8& bb0&{\small bb0} $\rightarrow$ {\small bb0} &  3 & $ b^2 d $ 
     & 0 & $\frac{1+2g}{3}$ & 0 & $\frac{2-2g}{3}$   \\
     9& bb1&{\small bb1} $\rightarrow$ {\small bb0} &  3 & $ b^2 u $  
     & 0 & $\frac{1+2g}{3}$ & 0 & $\frac{2-2g}{3}$   \\
10& bbb&{\small bbb} $\rightarrow$ {\small bbb} &  1 & $ b^3 $  
     & 0 & $g$ & 0 & $1-g$ \\
     [0.7ex]
      \cline{2-3} \\% \cmidrule(l){2-3}\\
     [-2.5ex]
11& a00&{\small a00}  $\rightarrow$ {\small a00} &  3 & $ a d^2 $  
     & 0 & $\frac{2-f}{3}$ & $\frac{1+f}{3}$ & 0 \\
12& a10&{\small a10}  $\rightarrow$ {\small a11} &  6 & $ a u d $  
     & $\frac{2+g}{3}$ & 0 & $\frac{1-g}{3}$ & 0 \\
13& a11&{\small a11}  $\rightarrow$ {\small a11} &  3 & $ a u^2 $  
     & $\frac{2+g}{3}$ & 0 & $\frac{1-g}{3}$ & 0  \\
     [1.0ex]
      \cline{2-3} \\% \cmidrule(l){2-3}\\
     [-2.5ex]
14& ab0&{\small ab0} $\rightarrow$ {\small ab0}  &  6 & $ a b d $  
     & 0 & $\frac{1-f+g}{3}$ & $\frac{1+f}{3}$ & $\frac{1-g}{3}$  \\
15& ab1&{\small ab1} $\rightarrow$ {\small ab1}  &  6 & $ a b u $  
     & $\frac{1+g-f}{3}$ & 0 & $\frac{1-g}{3}$ & $\frac{1+f}{3}$  \\
16& abb&{\small abb} $\rightarrow$ {\small abb}  &  3 & $ a b^2 $  
     & 0 & $\frac{-f+2g}{3}$ & $\frac{1+f}{3}$ & $\frac{2-2g}{3}$  \\
     [1.0ex]
17& aa0&{\small aa0} $\rightarrow$ {\small aa1} &  3 & $ a^2 d $  
     & $\frac{1+2g}{3}$ & 0 & $\frac{2-2g}{3}$ & 0  \\
18& aa1&{\small aa1} $\rightarrow$ {\small aa1} &  3 & $a^2 u$  
     & $\frac{1+2g}{3}$ & 0 & $\frac{2-2g}{3}$ & 0  \\
19& aab&{\small aab} $\rightarrow$ {\small aab}  & 3 & $ a^2 b $  
     & $\frac{2g-f}{3}$ & $0$ & $\frac{2-2g}{3}$ & $\frac{1+f}{3}$  \\
     [1.0ex]
20& aaa&{\small aaa}  $\rightarrow$ {\small aaa} & 1 & $ a^3 $  
     & $g$ & 0 & $1-g$ & 0  \\
     [0.7ex]
      \hline % \bottomrule
\end{tabular}
%v5 (2.7
\caption{The $r=3$ group agent pattern table for the system with floaters (0/1) and inflexibles (a/b).  See the main text for the explanation of the items.
In the fifth column, $p$ is the ratio of agents supporting $A$ party,  and $a$ and $b$ are the ratio of $A$- and $B$-inflexibles among all agents, and  $u$, $d$ are defined by $u=p-a$, $d=1-p-b$.
%The first column is an indexing label, the second column, all possible formation patterns of agents in a group of size $r=3$ disregarding the order of appearance, and the third column, the binary value of the agents in that group before and after the update.  The forth column is the multiplicity of the pattern coming from the different orderings, and the fifth column, the probability of the occurrence of the pattern.  Sixth, seventh, eighth, and ninth columns are the probability of obtaining an agent with opinion $1$ after the update.and the sixth column, 
}
\label{tab:booktabs}
\end{table}
%--------------------
%
%%
%\begin{eqnarray}
%\label{}
%\end{eqnarray}
%%
%\begin{figure}[htbp]

%At $N \to \infty$ limit. 
%v5 (2.6
The evolution of $\{ p_t, a_t, b_t \}$ can be calculated by tabulating all possible agent configurations of a group (indexed by $k$), their multiplicity ($m_k$), the probability of their occurrence ($P_k$), their contributions to the appearance of each agent type and values ($K^{f1}_k$ and $K^{f0}_k$ for floaters supporting $A$  and $B$ parties respectively, and $K^a_k$ and $K^b_k$ for $A$ and $B$ inflexibles), taking their products and summing by $k$  in the form,
% 
%v5 (4.4
\begin{eqnarray}
\label{e101}
p_{t+1} = \sum_k m_k  P_k(p_t,a_t,b_t) (K_k^{f1} + K_k^{a})
\end{eqnarray}
\begin{eqnarray}
\label{e102}
a_{t+1} = \sum_k m_k  P_k(p_t,a_t,b_t) K_k^{a}
\end{eqnarray}
\begin{eqnarray}
\label{e103}
b_{t+1} = \sum_k m_k  P_k(p_t,a_t,b_t) K_k^{b}
\end{eqnarray}
For $r=3$, we have tabulated these quantities explicitly in Table 1, from which, we obtain the formulae for the evolution of probabilities in the form of three variable dynamical difference equation as,
%%
%% post publication error correction apr 20, 2018
%critical typo!!
%
\begin{eqnarray}
\label{e111}
\!\!\!\!\!\!\!\!\!\!\!\!\!\!\!\!\!\!\!\!%%
p_{t+1} = -2p_t^3+3p_t^2+(1+f)\left[ (1-p_t)^2 a_t -p^2 b_t \right]
\nonumber \\
+\frac{1}{3}h \left[ (1-p_t-b_t)^3 -(p_t-a_t)^3 \right] \ ,
\end{eqnarray}
\begin{eqnarray}
\label{e112}
a_{t+1} = a_t \left\{ 1-g+ (f+g)(1-p_t)^2 \right\} + \frac{1}{3} h (1-p_t-b_t)^3 \ ,
\end{eqnarray}
%%
%v5 (2.8
\begin{eqnarray}
\label{e113}
\!\!\!\!\!\!\!\!\!\!\!\!\!\!\!\!\!\!\!\!\!\!\!\!\!\!\!\!\!\!%%
b_{t+1} = b_t \left\{ 1-g+ (f+g)p_t^2 \right\} + \frac{1}{3} h (p_t-a_t)^3 \ ,
\end{eqnarray}
under the physical constraint $0\leq a_t \leq p_t  \leq1-b_t \leq 1$.  These are the equation, whose solutions we shall analyze in following sections.

When $f=g=h=0$, ratios of inflexibles become constant as $a_t=a$ and $b_t=b$ reducing above set of equations to,
\begin{eqnarray}
\label{e11c}
p_{t+1} = a-2 a p_t + (3+a-b) p_t^2 - 2 p_t^3 ,
\end{eqnarray}
which is the evolution equation for normal $r=3$ Galam model.  It has been established that the evolution of $p_t$ described by (\ref{e11c}) has two attractive fixed points {\em or} a single attractive fixed point depending on the value of $a$ and $b$ \cite{GA05a, GJ07}.  For $b=0$, specifically, a critical  value $a^\star=3-2\sqrt{2} \approx 0.172$ separates these two cases.  At $a$ infinitesimally smaller than $a^\star$, we have the two attractive fixed points and a separator at $p^\star=1-\frac{1}{\sqrt{2}}$, which delimits $A$-winning and $B$-winning initial values of $p$. 

%%%%
%%%%
\section{Numerical analysis}

%
%{\SG2 I have reorganized that part putting first the Figures description }
%
%%%%%%%
%
\begin{figure}[h]
\begin{center}
\includegraphics[width=12.1cm]{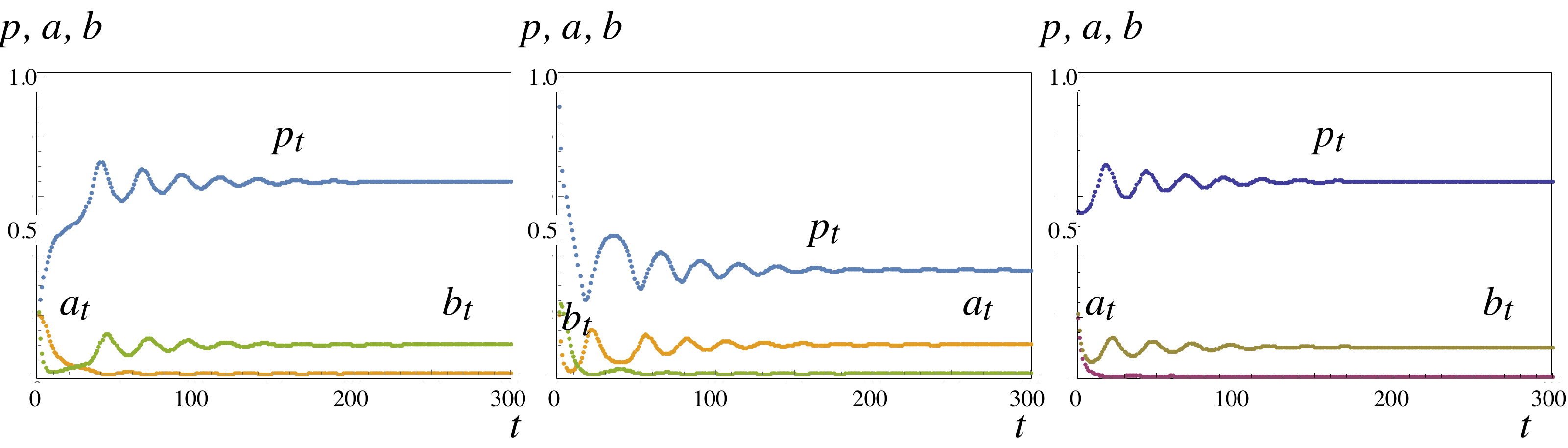}
\includegraphics[width=4.4cm]{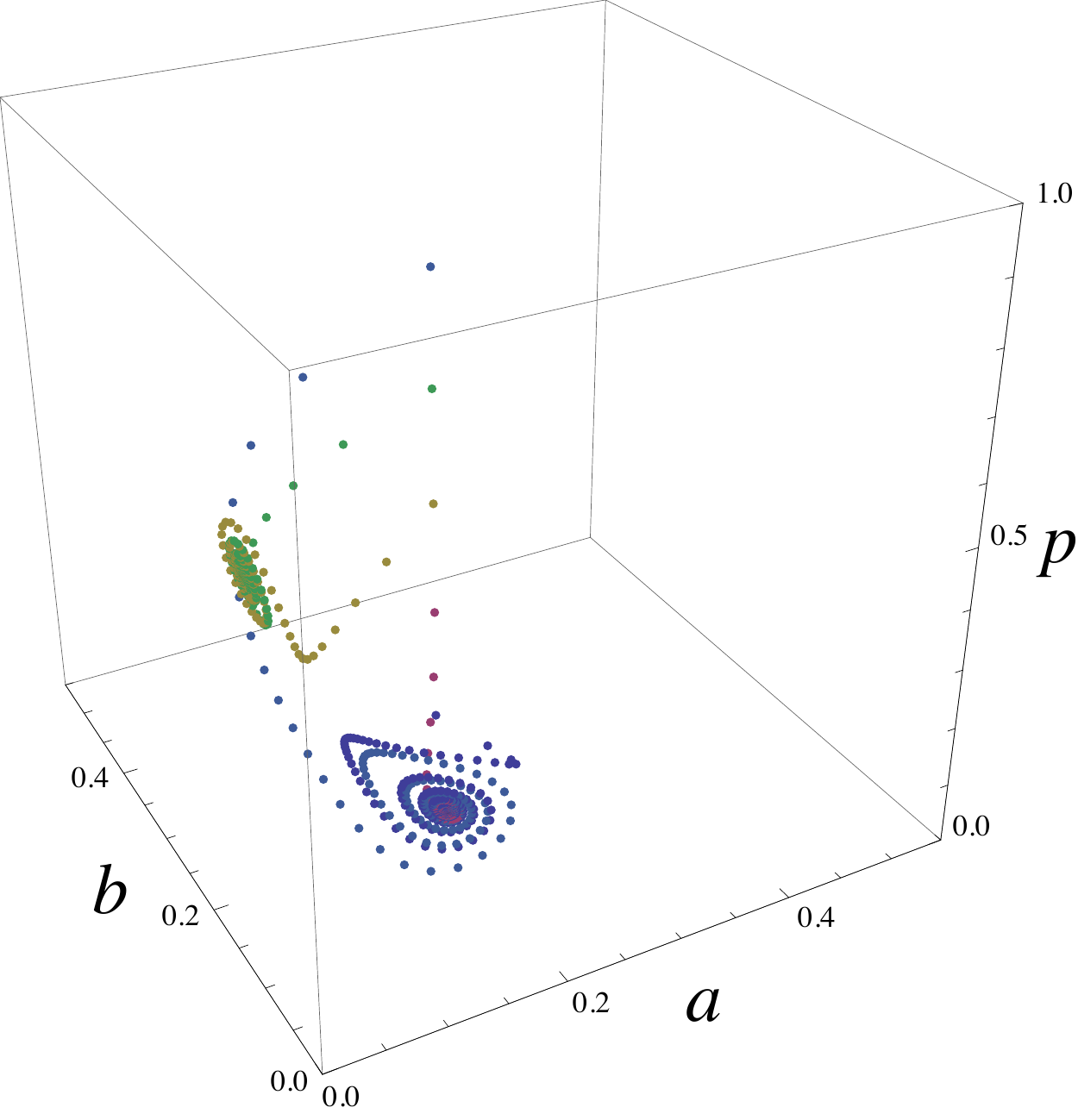}
\includegraphics[width=4.4cm]{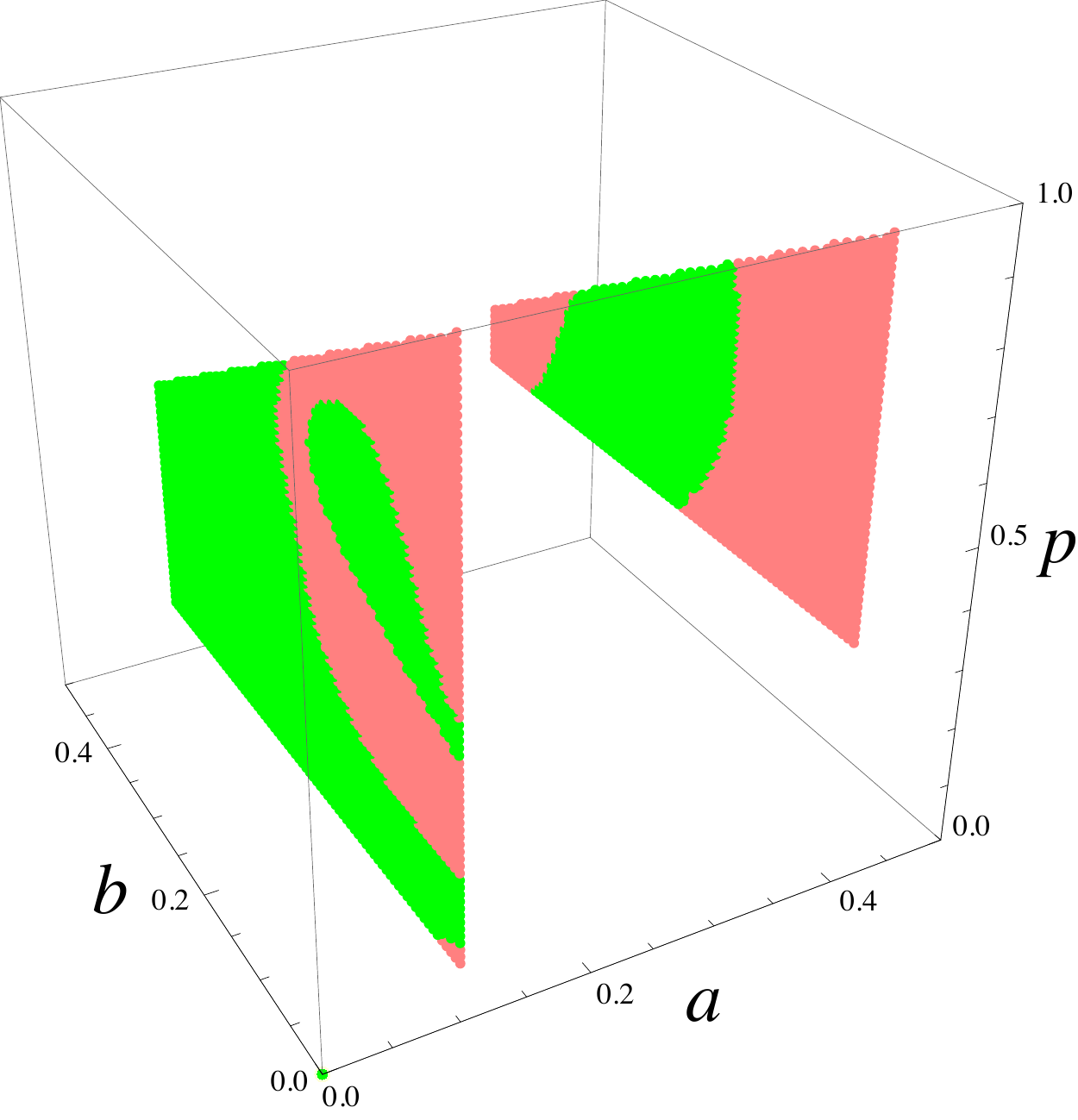}
\end{center}
%v5 (3.2 (3.3
\caption{%(Color online) 
The top graph represents the temporal evolutions of probabilities $p_t$ (blue), $a_t$ (orange) and $b_t$ (green) starting from three different initial conditions. The bottom-left graph depicts the phase space trajectory $\{ p_t, a_t. b_t\}$.  Different colors indicate trajectories of different initial conditions.  Two attractors around $\{ p, a. b\} \approx \{ 0.65, 0, 0.15 \}$ ($A$-dominant attractor) and $\{ p, a. b\} \approx \{ 0.35, 0.15, 0 \}$  ($B$-dominant attractor)are clearly visible as the center of spiral orbits.  The bottom-right graph depicts two sections of basin of attraction: red for the basin of $A$-dominant attractor, green for the basin of $B$-dominant attractor.
Parameters of the system are $f=0.1$, $g=0.45$, $h=0.25$.}
\label{fg1}
\end{figure}
%
%%
%\begin{figure}[htbp]
%\begin{figure}[h]
%\label{fg3}
%\begin{center}
%\includegraphics[width=12cm]{f22.pdf}
%
%\includegraphics[width=4.2cm]{f21.pdf}
%\includegraphics[width=4.2cm]{f23.pdf}
%\end{center}
%\caption{(Color online)  The temporal evolutions of probabilities $p$, $a$, $b$ (top), and phase space 
%trajectory (bottom left), and basin of attraction (bottom right).
%Parameters of the system are chosen as $f=0.1$, $g=-0.37$, $h=0.20$.}
%\end{figure}
%
%\begin{figure}[htbp]
\begin{figure}[h]
\begin{center}
\includegraphics[width=12.1cm]{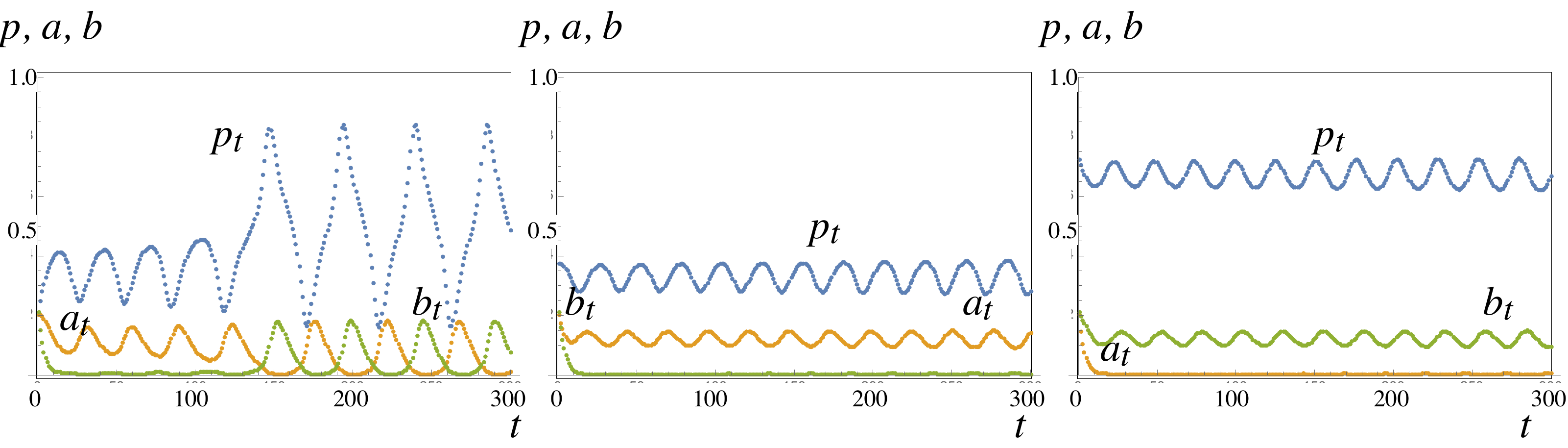}
\includegraphics[width=4.4cm]{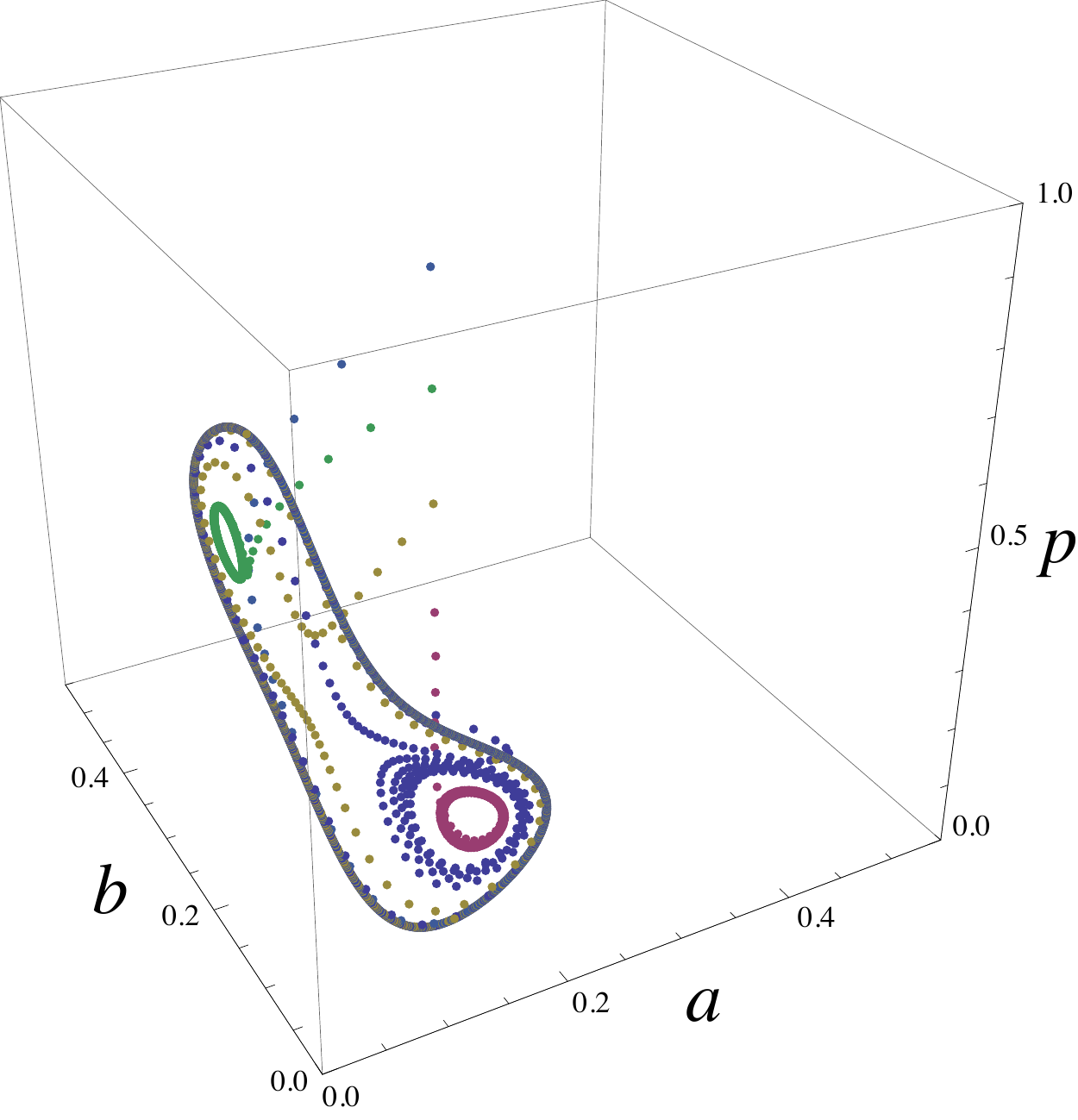}
\includegraphics[width=4.4cm]{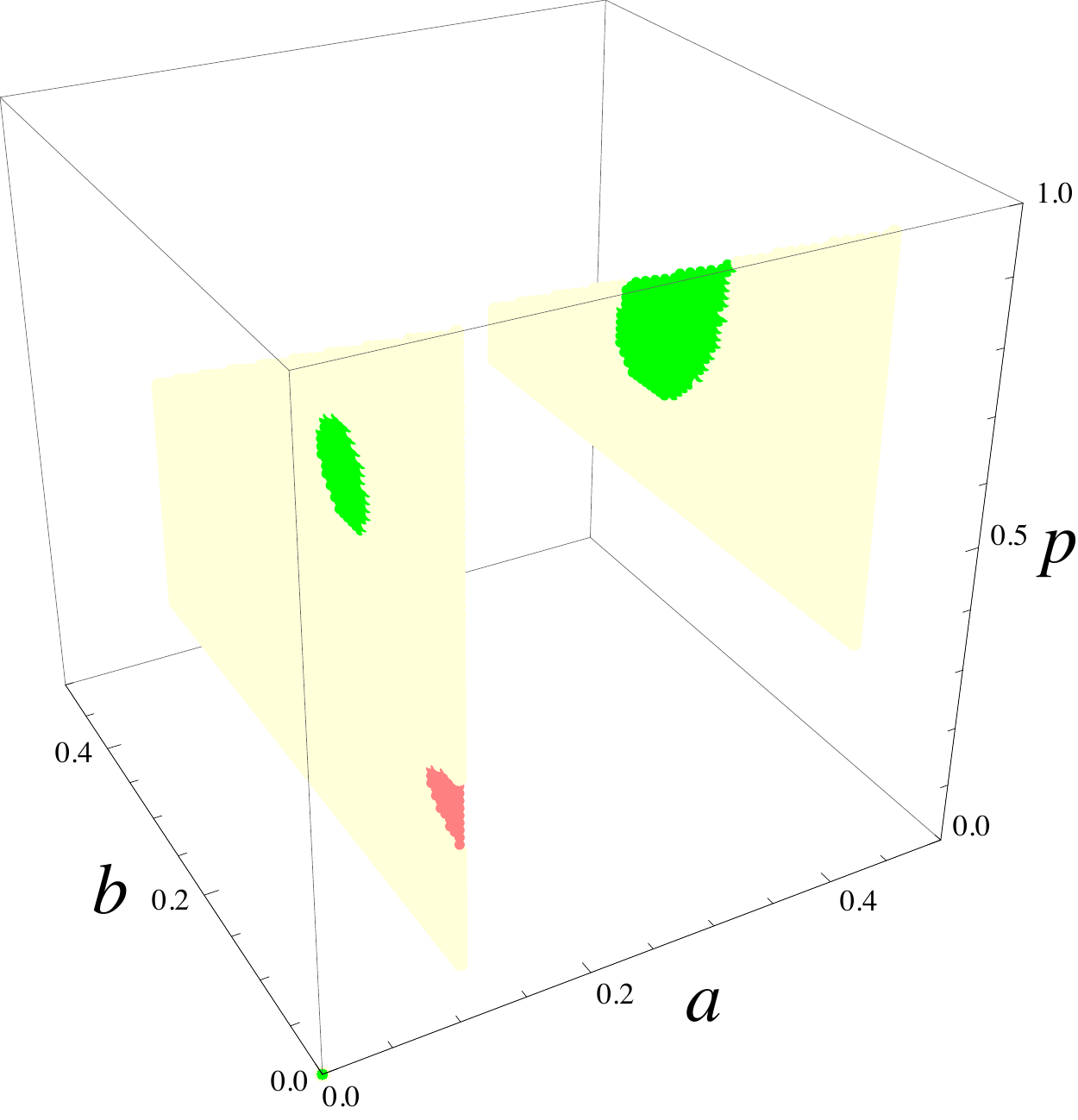}
\end{center}
%v5 (3.6
\caption{%(Color online)  
The temporal evolutions of probabilities $p$, $a$, $b$ (top), the phase space trajectory (bottom left), and two sections of basin of attraction (bottom right; red for $A$-attractor, green for $B$-attractor, yellow for $C$-attractor).
Parameters of the system are $f=0.1$, $g=0.31$, $h=0.15$.}
\label{fg2}
\end{figure}
%
%%
%
%\begin{figure}[htbp]
\begin{figure}[h]
\begin{center}
\includegraphics[width=12cm]{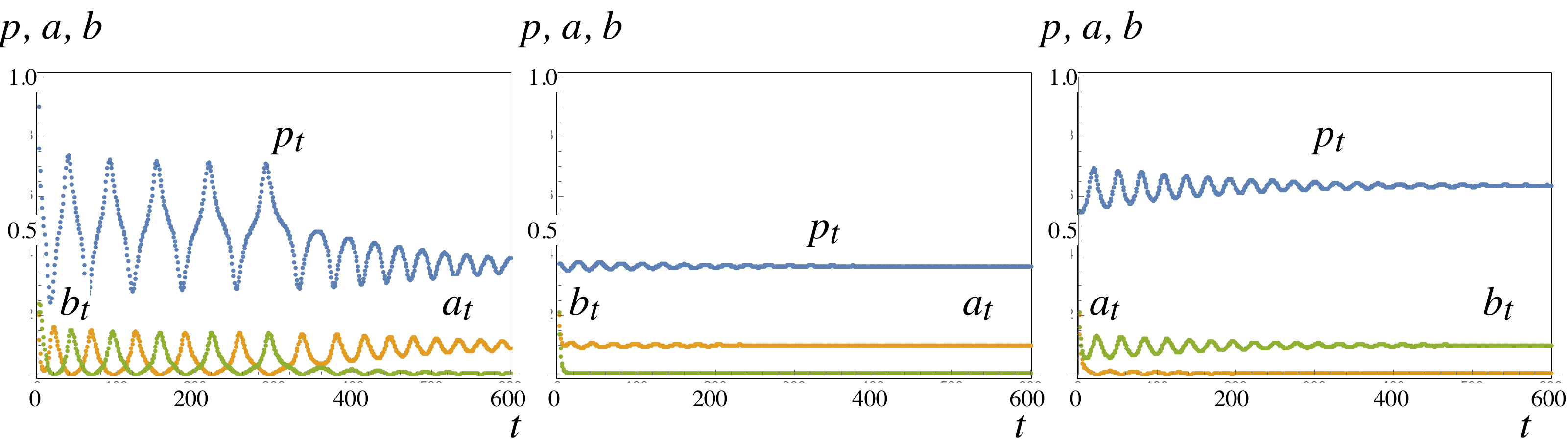}
\includegraphics[width=4.2cm]{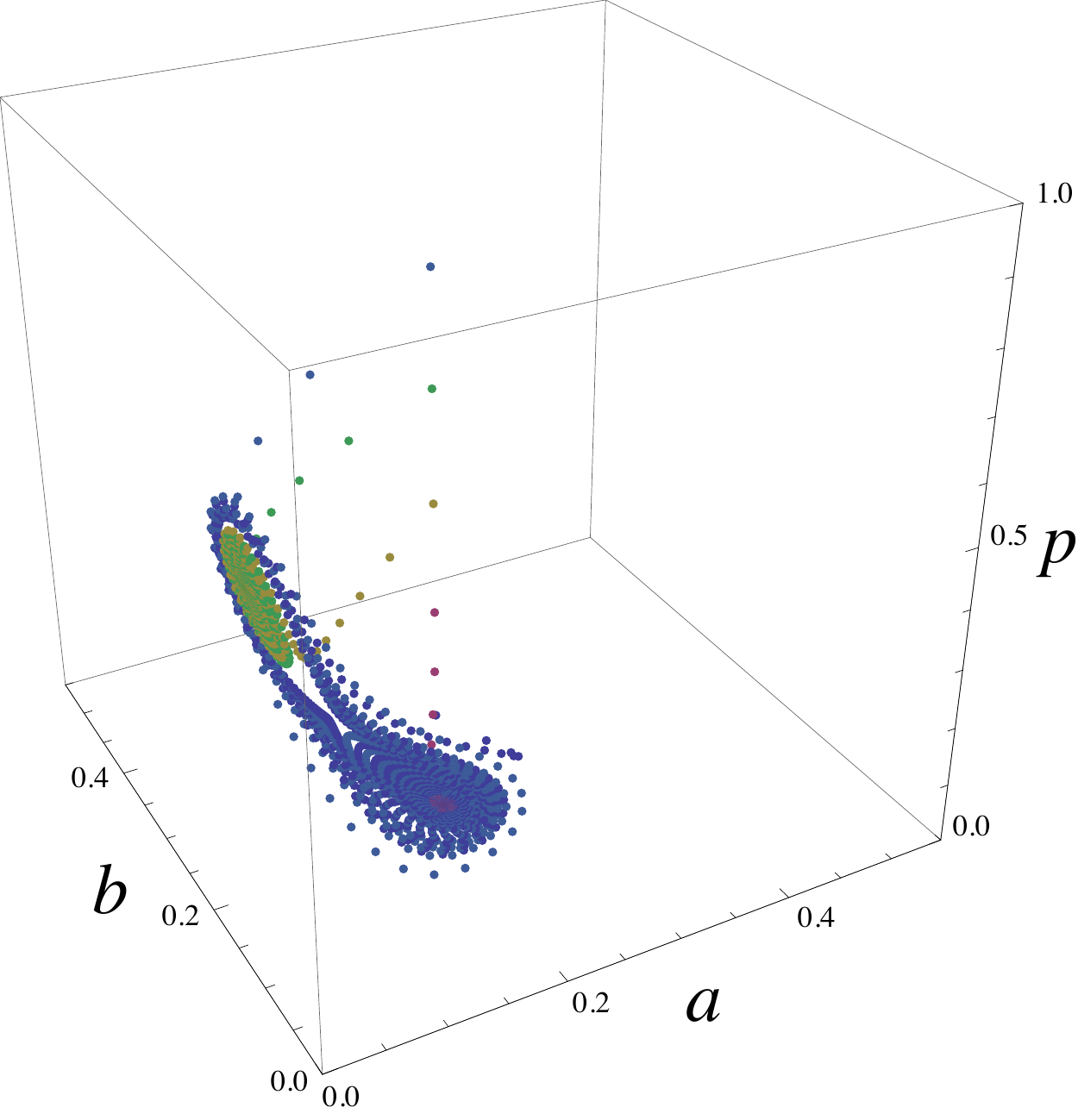}
\includegraphics[width=4.2cm]{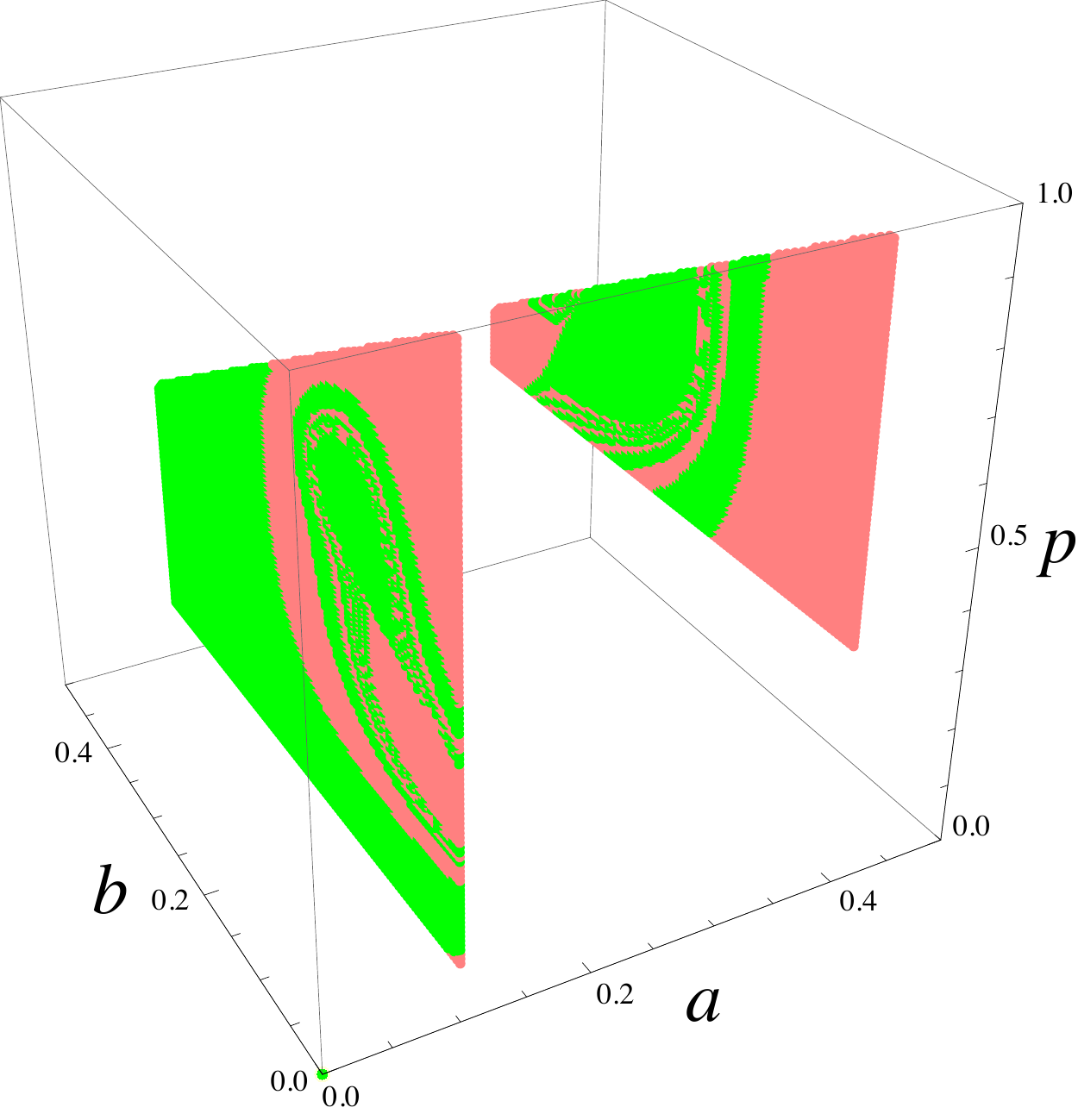}
\end{center}
\caption{%(Color online)  
The temporal evolutions of probabilities $p$, $a$, $b$ (top), and phase space rajectory (bottom left), and two sections of basin of attraction (bottom right; red for $A$-attractor, green for $B$-attractor).
Parameters of the system are $f=0.1$, $g=0.421$, $h=0.25$.}
\label{fg3}
\end{figure}
%

%v5 (3.1
To illustrate the dynamics driven by Eqs. (\ref{e111}), (\ref{e112}) and (\ref{e113})  some typical numerical examples of the evolution of probabilities, $p_t$, $a_t$ and $b_t$, along with phase space trajectories $\{ p_t, a_t, b_t\}$ are exhibited in Figures \ref{fg1}, \ref{fg2}, and \ref{fg3}.  

Figure \ref{fg1} represents the results for the parameter set $\{ f, g, h\} = \{ 0.1, 0.45, 0.25\}$, for which all initial states lead to one of two fixed points through inwardly spiraling trajectories.  In this case there is always a clear-cut winner party for all initial states.
%v5 R4
The damped oscillation of probabilities $p_t$, $a_t$ and $b_t$ can be seen as the reflection of the spiraling convergence to the fixed points in the phase space. 
%

%v5 R4
Figure \ref{fg2} represents the results for the parameter set $\{ f, g, h\} = \{ 0.1, 0.31, 0.15\}$, for which all trajectories lead to one of three limit cycles, which are revealed in $p_t$, $a_t$ and $b_t$ as the oscillating behavior.
%v4 R3 1
There are possibilities of either party winning the contest or of both parties winning the contest in alternating fashion.  Note that the basin of attraction looks larger for $C$-attractor basin, signifying the dominance of regime-changing cycle.  With the slight increase of $f$, or the slight decrease of $g$, both the $A$- and $B$-attractor basins disappear completely, and all initial states lead to regime-changing limit cycle.
%
%three type of trajectory can be observed.  There are trajectories leading to one of two attractors corresponding to $A$ or $B$ party dominance, just as in Figure 1.  Then, there are trajectories which settles into the limit cycle that oscillates between the party $A$-dominance  and the party $B$ dominance.  These three trajectories coexist in the phase space, as the basin of attraction for each 
%three final states are painted in different colors.
%
%
%There is a case of coexistence of three limits cycles, which we show in Figure 3, with the parameter set $\{ f, g, h\} = \{ 0.1, -0.31, 0.15\}$.   One limit cycle is the analogue of previous case signifying the oscillatory regime change, and other two representing one party dominance with cyclic ebb and flow of extremists in minority group, which can be thought of as representing the infighting for control.

Figure \ref{fg3} is an interesting ``borderline'' case between two previous examples with the parameter set $\{ f, g, h\} = \{ 0.1, 0.421, 0.25\}$.  There, all initial states lead to one of two fixed points, but often barely, after long period of wandering around limit cycle-like oscillation.  The basin of attraction shows very complex patterns reminiscent of dynamics at {\em the critical point of phase transition}.  It appears that  this case delimits the two phases shown in Figures  \ref{fg1}, \ref{fg2} respectively representing the majority-preserving attractor dominance and the regime-changing attractor dominance.   We shall see in the following section, that this hypothesis of phase transition is corroborated by further analysis.

%v5 (3.4
From these figures we can recognize that all trajectories eventually approach either to one of two fixed points, or to one of three limit cycles.  
%v5 (3.5
Two fixed points are symmetric to each other with respect to the transformation $\{ p, a, b\} \leftrightarrow \{ 1-p, b, a\} $, each representing the stable $A$-majority ($p>\frac{1}{2}$) and $B$-majority ($p<\frac{1}{2}$) states.  Two of the limit cycles are also symmetric to each other with respect to the same transformation 
%$\{ p, a, b\} \leftrightarrow \{ 1-p, b, a\} $, 
each encircling one of the fixed points, representing the oscillating $A$-majority ($p>\frac{1}{2}$) and $B$-majority ($p<\frac{1}{2}$) states.  

%v5 R3 1
It is notable that fixed points and limit cycle surrounding them have the roughly two-to-one split for the suppoters of winning and losing parties, namely  $p:(1-p) \approx 2:1$.
The third limit cycle which circles around both fixed points differs from the above two limit cycles in its majority-alternating nature.  It represents periodic regime-changing oscillation.  It is interesting to observe that the period of oscillation for the majority-alternating cycle appears to be twice that for the majority-preserving cycles.

Bunching together the fixed point and the majority-preserving limit cycle encircling it, we can classify the final states of the system into three; $A$-majority ($A$), $B$-majority ($B$), and regime-changing ($C$) attractors.   In the Figures, basins of attraction for each of these three attractors are drawn with color codes:  red for $A$-attractor basin, green for $B$-attractor basin and yellow for $C$-attractor basin.

%%%%%%%

%%%%
\section{Fixed points and their stability}
% fixp %

%v5 (4.1
We now look at the fixed points of the system \cite{AS96} in order to gain insight of the dynamics we have observed in last section.
For  the general case, it is hard to obtain the explicit expressions for the fixed point of our system.   However, we can obtain a good insight by examining the special case of $h=0$, for which three sets of fixed points can be analytically written down: 
\begin{eqnarray}
\label{e121}
p^{\star(0)}  = \frac{1}{2},
\quad
a^{\star(0)}  = 0 ,
\quad
b^{\star(0)}  =0 .
\end{eqnarray}
%%
%
%v5 R4
\begin{eqnarray}
\label{e131}
p^{\star(1)}  = \sqrt{\frac{g}{g+f}},
\quad
a^{\star(1)}  = 0 ,
\quad
b^{\star(1)}  = \frac{1}{1+f} \left\{  3-\frac{f+3g}{g} p^\star_1 \right\}
\end{eqnarray}
\begin{eqnarray}
\label{e132}
p^{\star(2)}  = 1-\sqrt{\frac{g}{g+f}},
\quad
a^{\star(2)}  = \frac{1}{1+f} \left\{  3+\frac{f+3g}{g} p^\star_1 \right\} ,
\quad
b^{\star(2)}  = 0 .
\end{eqnarray}
Note that with a special choice $g=f$ we have,
\begin{eqnarray}
\label{e141}
p^{\star(1)}  = \frac{1}{\sqrt{2}},
\quad
a^{\star(1)}  = 0 ,
\quad
b^{\star(1)} = \frac{3-2\sqrt{2}}{1+f} .
\end{eqnarray}
\begin{eqnarray}
\label{e142}
p^{\star(2)} = 1-\frac{1}{\sqrt{2}},
\quad
a^{\star(2)} =  \frac{3-2\sqrt{2}}{1+f} ,
\quad
b^{\star(2)} = 0.
\end{eqnarray}
%
%v5 (4.2
A notable fact is that the critical separating point $3-2\sqrt{2}\approx 0.172$ of original Galam model with $r=3$ shows up in the expression of our critical points $a^\star$ and $b^\star$, and it  is associated with the critical point $p^\star = \frac{1}{\sqrt{2}} \approx 0.707$ \cite{GJ07}.   
% 7-3 splits %
%v5 (4.3
These numbers -- two-to-one majority-minority splits with the minority comprising  some 60 \% inflexibles -- are the key to the understanding of the dynamics of our model system.

%v5 (4.5
The characteristics of the temporal evolution of the system around the fixed point can be analyzed using the linearized equations.  From the definition $\delta p_t=p_t - p^\star$, $\delta a_t=a_t - a^\star$, $\delta b_t=b_t - b^\star$, taking only their linear terms, we obtain the linear map,
\begin{eqnarray}
\begin{pmatrix} \delta p_{t+1} \\ \delta a_{t+1} \\ \delta b_{t+1} \end{pmatrix}
= M \begin{pmatrix} \delta p_{t} \\ \delta a_{t} \\ \delta b_{t} \end{pmatrix} ,
\end{eqnarray}
where $M$ is the stability matrix of dimension $3 \time 3$.
%v5 (4.6
For the fixed point  $\{ p_0^\star, a_0^\star, b_0^\star\}$,  we get the stability matrix $M^{(0)}$ from a simple calculation,
\begin{eqnarray}
M^{(0)}  = 
\begin{pmatrix} \frac{3}{2} & \frac{1+f}{4} & -\frac{1+f}{4} \\
0 & \frac{4+f-3g}{4} & 0 \\ 
0 & 0 & \frac{4+3f-g}{4} \end{pmatrix} .
\end{eqnarray}
Its eigenvalues,
\begin{eqnarray}
\begin{pmatrix} \lambda^{(0)}_1 \\  \lambda^{(0)}_2 \\ \lambda^{(0)}_3 \end{pmatrix}
= \begin{pmatrix} \frac{3}{2} \\  \frac{4+f-3g}{4} \\ \frac{4+3f-g}{4} \end{pmatrix},
\end{eqnarray}
one of whom is always lager than $1$, shows that this fixed point is unstable. 
% In fact, it is easy to see that this fixed point represent the separator.
%

For the fixed points  $\{ p_1^\star, a_1^\star, b_1^\star\}$ and $\{ p_2^\star, a_2^\star, b_2^\star\}$, it is enough to analyze just one of them, since the other is simply a mirror image $\{p, a, b\} \leftrightarrow \{1-p, b, a\}$, and has identical characteristics.
With a straightforward calculation we obtain the stability matrix $M^{(1)}$  for $\{ p_1^\star, a_1^\star, b_1^\star\}$ in the form,
\begin{eqnarray}
M^{(1)}  = 
\begin{pmatrix} \frac{2f}{f+g} 
& \!\!\!\!\!\!\!\!\!\!\!\!\!\!\!\! (1+f)\left(\frac{f+2g}{f+g} -2p^{\star(1)} \right)  
&\!\!\!\!  -(1+f) (p^{\star(1)})^2\\
0 
& \!\!\!\!\!\!\!\!\!\!\!\!\!\!\!\! (f+g+1)-2(f+g)p^{\star(1)} 
&\!\!\!\! 0 \\ 
 \frac{2(f+g+1)-4(f+g)p^{\star(1)}}{1+f} 
& \!\!\!\!\!\!\!\!\!\!\!\!\!\!\!\! 0 
&\!\!\!\! f-g+1 \end{pmatrix} .
\end{eqnarray}
%
%v5 (4.7
With eigenvalues, 
\begin{eqnarray}
\begin{pmatrix} \lambda^{(1)}_1 \\  \lambda^{(1)}_2 \\ \lambda^{(1)}_3 \end{pmatrix}
= \begin{pmatrix} f+g+1-2(f+g)p^{\star(1)}\\  \frac{f-g}{2}+\frac{3f+g}{2(f+g)}-\frac{ \sqrt{D} }{2} \\ \frac{f-g}{2}+\frac{3f+g}{2(f+g)}+\frac{ \sqrt{D} }{2} \end{pmatrix} ,
\label{leigen}
\end{eqnarray}
where we set,
%
%v5 (4.8
\begin{eqnarray}
D = (f-g-1)^2+4g+4 p^{\star(1)} (p^{\star(1)}-1) \{ (p^{\star(1)})^2+p^{\star(1)}-6g\},
\end{eqnarray}
which determines the stability of the orbits around the fixed point.  
%
%\begin{figure}[htbp]
\begin{figure}[h]
\label{fg4}
\begin{center}
\includegraphics[width=12cm]{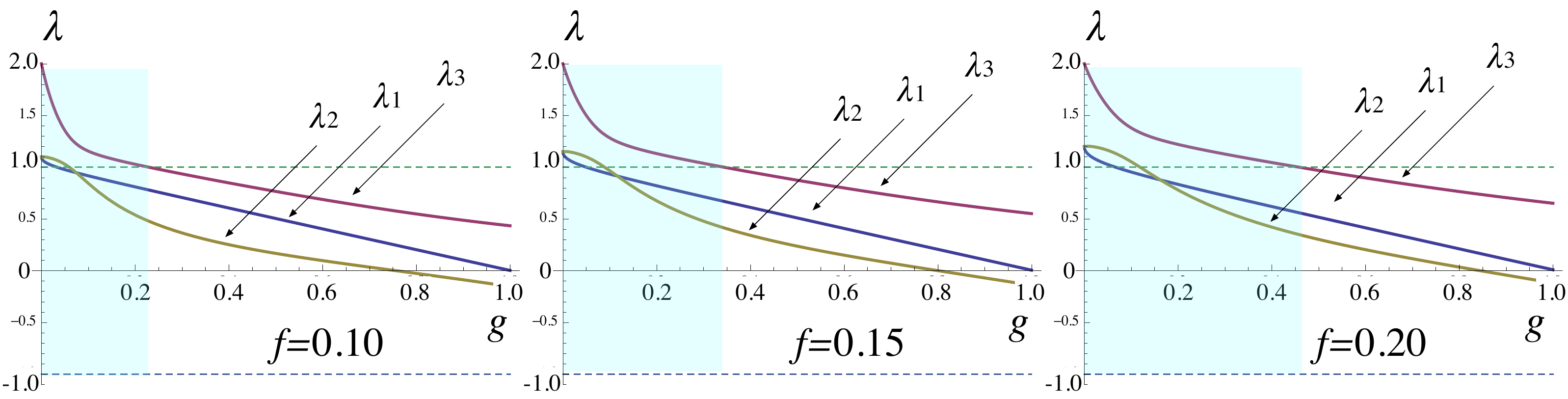}
\end{center}
%v5 (4.9
\caption{Eigenvalues of stability matrix $M^{(1)}$ for $h=0$ as the function of $g$ with fixed $f$.  Left $f=0.1$, and center $f=0.15$, right $f=0.2$.  The water color indicates the region where the fixed points are unstable.}
\end{figure}
%
%
%\begin{figure}[htbp]
\begin{figure}[h]
\begin{center}
\includegraphics[width=12cm]{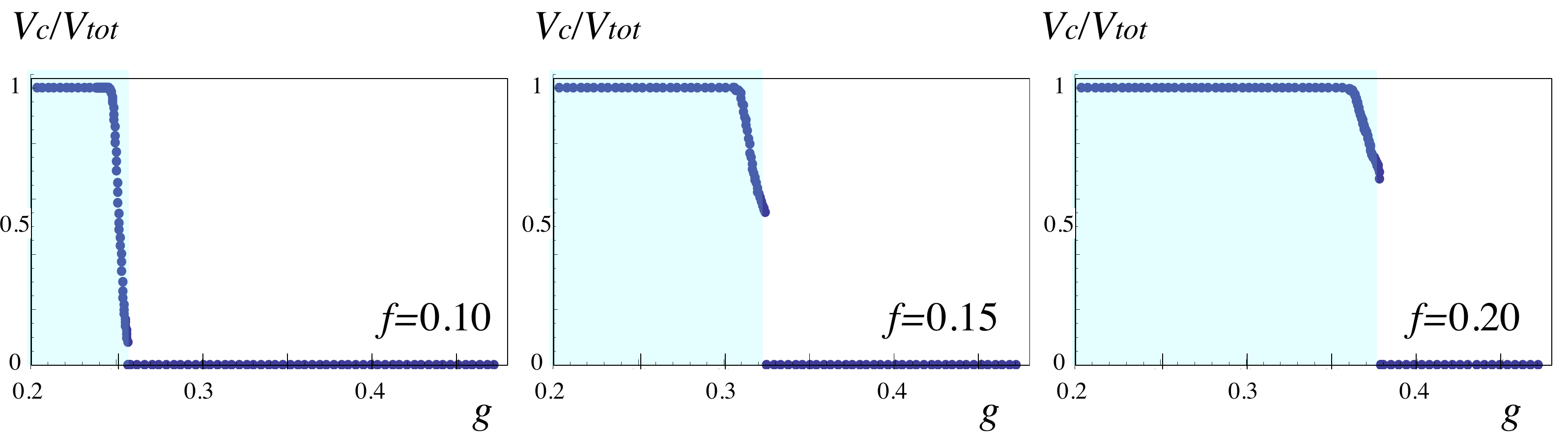}
\end{center}
\caption{Volume ratios of basin of attraction for majority-alternating limit cycle as a function of $g$ with fixed $f$ and $h$.  From left to right, $f=0.1$, $f=0.15$,  $f=0.2$ with fixed value $h=0.1$. The water color indicates the region where the final state is the regime-changing limit cycle.}
\label{fg5}
\end{figure}
%
%v5 (4.10
In Figure \ref{fg4} we show examples of the three eigenvalue of the stability matrix (\ref{leigen}) as functions of system parameter $g$ with fixed value of $f$.  
The fixed point is a spiral attractor where all three eigenvalues are less than $1$, while it is an unstable fixed point otherwise.  We can observe, in Figure \ref{fg5}, that the fixed points are stable attractors for larger $g$ region (white region) which is separated by a critical value of $g$ from the regions of unstable fixed points at smaller $g$ (water-colored region).  The increase of $f$ results in the increase of critical value of $g$. 

The numerical analysis of the system with $h\ne 0$, shown in Figure  \ref{fg5}, validates this analysis reasonably well.  There, the ratio of the volume in phase space $\{ p, a, b \}$ of the basin of attraction for the regime-changing attractor versus total volume is plotted as a function of $g$ for fixed values of $f$ and $h$.  The value $1$ signifies that the system always ends up in regime-changing limit cycles, while the value $0$ signifies that it always comes to stable majority for one side or the other.   The phase transition-like sudden change between $0$ and $1$ at a critical value  of $g$ is observed as we vary the parameter $g$.  Increasing $f$ results in a larger critical value for $g$, at which we observe the transition from the dominance of majority-preserving trajectories to that of majority-alternating trajectories.  
A comparison between Figures  \ref{fg4} and  \ref{fg5} reveals that the results obtained using linear stability analysis and shown in Figure  \ref{fg4} are qualitatively acceptable, if not quantitatively accurate, in predicting the dynamics of the system.

%%%
%%%
\section{Analysis of oscillations}

Finally, we turn our attention to the period of the limit cycles.  It manifests itself as the oscillating behavior of probabilities $\{p, a, b \}$.   As  noted in the analysis of previous sections, the period of regime-changing limit cycle is about twice that of majority-preserving limit cycle 
%v5 (5.1
(See top figures in Figures \ref{fg2} and \ref{fg3}), the latter of which we simply call the period $T$ hereafter.  Numerical experimentation reveals that the period $T$ does strongly depends on the parameters $f$ and $h$, but is rather insensitive to change in $g$.  We take this fact into account in our following analysis.

%v5 (5.2 (made figs bigger
%\begin{figure}[htbp]
\begin{figure}[h]
\begin{center}
\includegraphics[width=3.2cm]{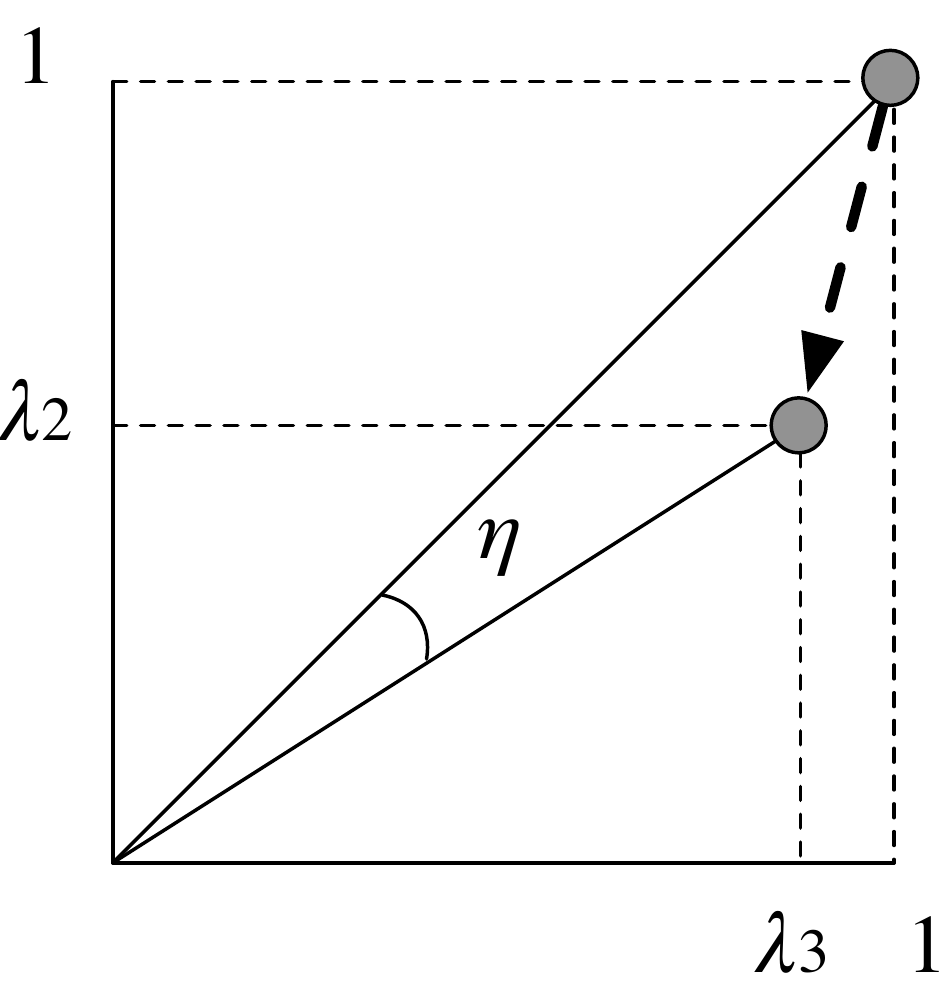}
\end{center}
\caption{The angle of rotation at each time step $\eta$ in the process of asymmetric shrinking  $\{1, 1\} \to \{\lambda^{(1)}_2, \lambda^{(1)}_3\}$.}
\label{fg6}
\end{figure}
The period of oscillation $T$ can be roughly estimated from the angle of rotation at each time step $\eta$ in the process of asymmetric shrinking (Figure \ref{fg6}), which is largest between $\lambda^{(1)}_3$ and $\lambda^{(1)}_2$.  Namely,  $\{1, 1\} \to \{\lambda^{(1)}_2, \lambda^{(1)}_3\}$ induces the angular rotation $\eta$ given by,
\begin{eqnarray}
\tan\left( \frac{\pi}{4}-\eta \right) = \frac{\lambda^{(1)}_2}{\lambda^{(1)}_3} .
\end{eqnarray}
From this Equation, we can obtain the period $T =\frac{2\pi}{\eta}$.  When $h=0$ and $g=f$, using the leading term of the expansion in $f$, we  obtain
$ \frac{\lambda^{(1)}_2}{\lambda^{(1)}_3} \approx \sqrt{3\sqrt{2}-4}\sqrt{f}$, and therefore, 
\begin{eqnarray}
\label{t-th}
T \approx \frac{2\pi}{\sqrt{3\sqrt{2}-4}} \frac{1}{\sqrt{f}} .
\end{eqnarray}
%
%\begin{figure}[htbp]
\begin{figure}[h]
\begin{center}
\includegraphics[width=4.7cm]{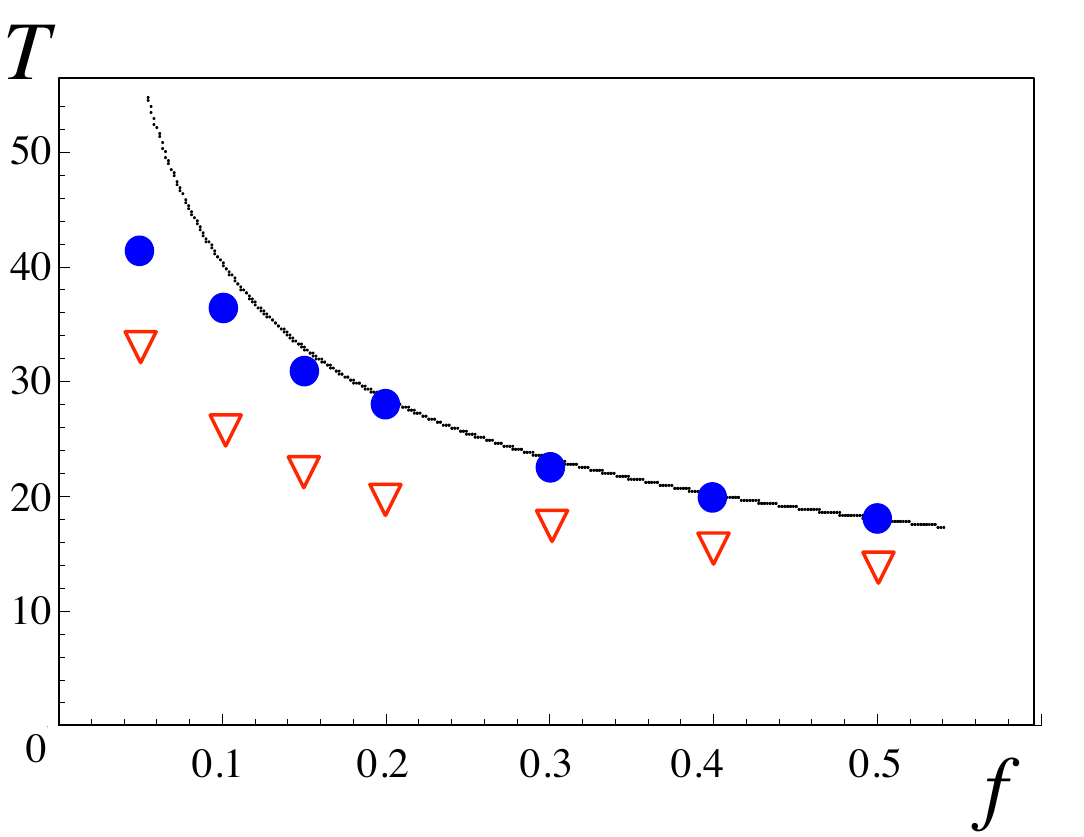}
\end{center}
%v4 (5.3
\caption{Theoretical estimate of the period of oscillation $T$ as a function of parameter $f$, with the numerical data obtained from $h=0.02$ (blue circle) and $h=0.05$ (red inverted triangle) with $g=f$.}
\label{fg7}
\end{figure}

In Figure \ref{fg7} we compare our prediction with the numerical data we have taken  $g=f$ for two values of $h$.  The agreement is surprisingly good for $h \to 0$, considering the roughness of the estimation solely based on the property of the fixed point.

\section{Summary}

In this paper, introducing an internal dynamical local dependence of inflexibility toward A or B as function of their respective majority/minority status, we have developed a model for shifting political opinions, formulated as an extension of the original Galam model for which the proportions of respective inflexibles are fixed external parameters.  

%v5 Rf2 2
It is possible to extend our model to the case of $r=4, 5, ...$ through straightforward but increasingly tedious calculations.  We have checked, by numerical calculations with $r=5$, that lager $r$ simply reproduces $r=3$ results with slightly shifted locations of fixed points and  limit cycles, and no substantial new features are observed. 

Because our model is a sufficiently complex nonlinear system, it is natural to expect it to yield not only fixed points, limit cycles and phase transition, but also strange attractors and chaotic dynamics.  Our numerical exploration of all segments of parameter space $\{ f, g, h\}$, however, has yielded no sign of chaos for now, 
We suspect that the introduction of some new process into the model seems to be necessary to bring about chaotic dynamics, which is an essential part of real-world political dynamics. 

%v5 R3 3
Although the current model, by itself is far from suitable to analyze actual data from real-world political arena, it seems to capture some basic features of evolution of political opinions.  Specifically, it is very intriguing that we find the existence of two different phases, majority-conserving and regime-changing, and sudden transition between them through the variation of system parameters. 
We hope our model will trigger further research for both its developments, refinements and eventual applications.          

\bigskip\bigskip
\noindent{\bf Acknowledgements}

We are grateful to Y. Matsuoka and D. Matsunaga for their assistance in the numerical calculations.
This research was supported by the Japan Ministry of Education, Culture, Sports, Science and Technology under the Grant number 15K05216.

\bigskip\bigskip
%\newpage
%%

%%

\end{document}